\documentclass[aps,prd,twocolumn,noeprint,showpacs,preprintnumbers,amsmath,amssymb]{revtex4-1}

\usepackage{graphicx}
\usepackage[usenames,dvipsnames]{xcolor}
\usepackage{braket}
\usepackage{amssymb}
\usepackage{amsfonts}
\usepackage{amsmath}
\usepackage{bbm} 
\usepackage[latin1]{inputenc} 
\usepackage{lineno}
\begin{document}
\title{Examples of symmetry-preserving truncations in tensor field theory}
\author{Yannick Meurice$^1$}
\affiliation{$^1$ Department of Physics and Astronomy, The University of Iowa, Iowa City, IA 52242, USA }
\definecolor{burnt}{cmyk}{0.2,0.8,1,0}
\def\lt{\lambda ^t}
\def\note{note}
\def\beq{\begin{equation}}
\def\enq{\end{equation}}
\newcommand{\Tr}{\text{Tr}}

\date{\today}
\begin{abstract}
We consider the tensor formulation of the non-linear O(2) sigma model and its gauged version (the compact Abelian Higgs model), on a $D$-dimensional cubic lattice, and show that tensorial truncations are compatible 
with the general identities derived from the symmetries of these models. This means that the universal properties of these models can be reproduced with highly simplified formulations desirable for implementations with quantum computers or for quantum simulations experiments. 
We discuss the extensions to global non-Abelian symmetries, discrete symmetries and pure gauge Abelian models.
\end{abstract}


\maketitle

\section{Introduction}
There has been a lot interest for tensorial formulations of lattice models in the context of the renormalization group method
\cite{nishino96,Levin2006,Gu:2009dr,PhysRevLett.103.160601,PhysRevB.86.045139,prb87,prd88,pre89,prd89,
PhysRevLett.115.180405,Shimizu:2014fsa,Shimizu:2014fsa,
Takeda:2014vwa,Shimizu:2017onf,PhysRevB.98.235148,Bal:2017mht,
Nakamura:2018enp,Kuramashi:2018mmi,Yoshimura:2017jpk,
Kadoh:2018hqq,juyu1,juysu2}. 
Tensor formulations provide a new approach of lattice models that we call tensor field theory (TFT). TFT should not be confused with theories involving fields that in the continuum have more than one Lorentz index, for instance the Kalb-Ramond field \cite{kalbramond}, and are often called ``tensor fields". 
For theories with compact fields like the nonlinear sigma models and Wilson lattice gauge theories, 
the tensor reformulation relies on character expansions and is always discretized \cite{prd88}. This is suitable for quantum computations or quantum simulations \cite{pra90,ahm,prl121}.
In practical situations such as Tensor Renormalization Group (TRG) calculations, truncations of infinite sums appearing in the TFT formulation of models with continuous symmetries are necessary. This can be achieved by discarding contributions to the partition function or observable averages that involve tensor indices larger than some cut-off value $n_{max}$. Concrete examples will be given in Secs. \ref{sec:o2} and 
\ref{sec:cahm}. 

A truncation procedure can be understood as a regularization and we need to ask if the regularization is  
compatible with the symmetries of the theory or if it generates what we call anomalies.   As far as the universal behavior is concerned, we expect that if truncations 
preserve the symmetries, one should be able to obtain  the properties associated with the universality classes by taking the contiunuum limit using a considerably simplified microscopic formulation. In other words, we could use drastic truncations of the sums such that at each site, link or plaquettes only a few values of the indices are kept. This is very important when the computational units available to represent the local degrees of freedom, such as qubits or 
trapped atoms, are in limited supply. 

In the following, we discuss identities associated with global and local symmetries in the Lagrangian approach of lattice models and examine their compatibility with truncations. 
We focus on two related examples with a continuous Abelian symmetry: the O(2) nonlinear sigma model and the compact Abelian Higgs model. We also connect with the Hamiltonian formulation by taking the time continuum limit. In the Hamiltonian approach, it is sufficient to check that the generators of symmetry groups commute with the Hamiltonian. We want to emphasize that the Lagrangian approach used in  the TRG and followed here is more general and that we will not rely on infinitesimal transformations as in the traditional Noether's approach.  The compatibility of the symmetries with truncations in TFT is a frequently asked question and we think that it is important to collect basic results about this question in situations where compact field integrations are replaced by discrete sums. 

The article is organized as follows. In Sec. \ref{sec:sym}, we introduce simple identities that are valid for global or local symmetries appearing in generic lattice models. 
In Sec. \ref{sec:o2}, we discuss the nonlinear O(2) sigma model in arbitrary dimension. This is an example of a model with a continuous global Abelian symmetry. In Sec. \ref{sec:cahm}, we consider the gauged version of the O(2) model, the compact Abelian Higgs model. In both cases, we find conclusive 
evidence that truncations fully preserve the symmetries of the model. Extensions to discrete symmetries, global non-Abelian symmetries and pure gauge Abelian theories are discussed in Sec. \ref{sec:ext}. In the conclusions, we summarize the results, provide an intuitive picture and emphasize the practical implications of the results.

\section{Implications of  symmetries for lattice models}
\label{sec:sym}

In this section, we consider a generic lattice model with action $S[\Phi]$, where $\Phi$ denotes a field configuration of fields $\phi_\ell$ 
attached to locations $\ell$ which can be sites, links, plaquettes or higher dimensional objects.  Additional indices possibly attached to the fields are kept implicit. 
The partition function reads
\beq
Z=\int {\mathcal D}\Phi {\rm e}^{-S[\Phi]}, 
\enq
with ${\mathcal D}\Phi$ the measure of integration over the fields. 
The average value of a function of the fields $f(\Phi)$ is defined as 
\beq
\langle f(\Phi) \rangle=\int {\mathcal D}\Phi f(\Phi){\rm e}^{-S[\Phi]}/Z.
\label{eq:sym}
\enq

We define symmetries as field transformations 
\beq
\phi_\ell\rightarrow\phi_\ell'= \phi_\ell +\delta \phi_\ell[\Phi], 
\enq
that 
preserve the action and the integration measure:
\beq 
{\mathcal D}\Phi' ={\mathcal D}\Phi\ {\rm and }\ S[\Phi']=S[\Phi].
\enq
These symmetries can be global or local. 
In all the examples we know, these symmetries form a group and the invariance is valid for any group element and not only for infinitesimal transformations. 
Changing variable from $\Phi$ to $\Phi'$ and using 
the symmetry properties of Eq. (\ref{eq:sym}), we find the intuitively clear result:
\beq
\langle f(\Phi) \rangle=\langle f(\Phi +\delta\Phi) \rangle.
\label{eq:inv}
\enq

Can this simple expression of the symmetries be used to derive the existence of conserved quantities 
for global continuous symmetries as in Noether's theorem? In classical mechanics, if a transformation $\delta q_i$ 
of generalized coordinates $q_i$ leaves the action invariant, then after using the equation of motion, we obtain conservation law:
\beq
\frac{d}{dt} (\frac{\partial L}{\partial \dot{q_i}}\delta q_i)=0.
\enq
The use of the equations of motion guarantees that the variation $\delta q_i$ has no effect {\it except} at the initial and final times where 
unlike what is done in the variational procedure $\delta q_i$  are not required to vanish. Consequently, 
the two individual surface terms do {\it not} vanish and are equal to the conserved quantity.  

In field theory, a similar procedure leads to 
a relativistically invariant current conservation 
\beq
\partial_\mu J^\mu(x)= \vec{\nabla }.\vec{J}+\partial \rho/\partial t=0,
\label{eq:cont}\enq which has the form of a continuity equation. By considering its integration between two time slices with spatial boundary conditions such that 
the spatial current does not flow outside the region of integration, one obtains that the integral of the charge density over a time slice is 
a constant of motion. 

In the following, we will show that Eq. (\ref{eq:inv}) can actually be obtained as a global consequence of a continuity equation encoded in the local tensors used in the reformulation. 
We will not need to use the equations of motion explicitly. In the generic formulation used above, the equations of motions are obtained by varying a single local variable $\phi_\ell$:
\beq
\phi_\ell\rightarrow\phi_\ell'= \phi_\ell +\alpha. 
\enq
Assuming that the ${\mathcal D}\Phi$ is invariant under this shift and that the action changes by an amount $\Delta_{\ell,\alpha}S$, we obtain that 
\beq
\langle {\rm e}^{-\Delta_{\ell,\alpha}S}\rangle=1.
\enq
Taking the derivative with respect to $\alpha$ and setting $\alpha=0$, we obtain the lattice equation of motion
\beq
\langle  \partial S/\partial \phi_\ell \rangle=0. 
\enq

\section {Example 1: the O(2) model}
\label{sec:o2}
\subsection{The model and its symmetry}
As a first example we consider a lattice model with a global continuous Abelian symmetry: the non linear O(2) sigma model.  
This is a generalization of the Ising model where the spins are two-dimensional vectors of length one. We parametrize them 
with an angle $\varphi$ where 0 and $2\pi$ are identified. We use a $D$-dimensional (hyper) cubic Euclidean space-time lattice. For instance, for $D=2$, we use a square lattice. 
The sites are denoted $x=(x_1, x_2,\dots x_D)$, with $x_D=\tau$, the Euclidean time direction. The total number of sites is denoted $V$ and we assume periodic or open boundary conditions. 
If we take the time continuum limit, we obtain a
quantum Hamiltonian formulation in 
$D-1$ spatial dimensions. 

In terms of the generic notations introduced in Sec. \ref{sec:sym}, the field configurations are 
$\Phi=\{\varphi_x\}_x$. The integration measure is normalized to one and reads
\beq
\int {\mathcal D}\Phi =\prod_x\int_{-\pi}^{\pi}\frac{d\varphi _x}{2\pi}, 
\enq
and the action
\beq
S[\Phi]=-\beta \sum\limits_{x,i} \cos(\varphi_{x+\hat{i}}-\varphi_x),\enq
where $\hat{i}$ denotes a unit vector in the positive $i$-th direction. 
The invariance requirements for the action and measure of Eq. (\ref{eq:sym}) are satisfied for the global shift
\beq
\varphi_x'=\varphi_x+\alpha. 
\label{eq:phivar}
\enq
This implies that for a function $f$ of $N$ variables
\beq
\langle  f(\varphi_{{x}_1},\dots ,\varphi_{{x}_N})\rangle = \langle  f(\varphi_{{x}_1}+\alpha, \dots ,\varphi_{{x}_N}+\alpha)\rangle. \enq
Since $f$ is $2\pi$-periodic  in its variables and can be expressed in terms Fourier modes, this can be reduced to 
\begin{eqnarray}
\label{eq:insert}
&\langle&  \exp(i(n_1\varphi_{{x}_1}+\dots n_N\varphi_{{x}_N}))\rangle =\\ \nonumber
&\ &\exp((n_1+\dots n_N)\alpha)\langle  \exp(i(n_1\varphi_{{x}_1}+\dots n_N\varphi_{{x}_N}))\rangle. 
\end{eqnarray}
This implies that if
\beq 
\label{eq:selection}\sum_{n=1}^N n_i \neq 0,\enq
 then \beq \langle  \exp(i(n_1\varphi_{{x}_1}+\dots +n_N\varphi_{{x}_N}))\rangle=0.\enq
We will show that this selection rule can be explained by a microscopic continuity equation that is manifest in the tensor formulation that 
we proceed to discuss. 
\subsection{The tensor formulation}
The basic aspects of the tensor reformulation of the O(2) model have been discussed in Refs. \cite{prd88,pre89,prd89}.
We briefly review the main results. It borrows tools from duality constructions \cite{RevModPhys.52.453}. At each link, 
we use the Fourier expansion\beq
            {\rm e}^{\beta  \cos(\varphi_{x+\hat{i}}-\varphi_x)} = \sum\limits_{n_{x,i}=-\infty}^{+\infty} {\rm e}^{i n_{x,i}(\varphi_{x+\hat{i}}-\varphi_x)} I_{n_{x,i}}(\beta)\  ,
            \label{eq:fou}
\enq
        where the $I_n$ are the modified Bessel functions of the first kind. This attaches an index $n_{x,i}$ at each link coming out of $x$ in the positive $i$-th direction.
It is then possible to integrate over the $\varphi_x$ and rewrite the partition function as the trace of a tensor product:
\beq
Z=I_0^V(\beta) {\rm Tr} \prod_x T^x_{(n_{x-\hat{1},1}, n_{x,1},\dots,n_{x,D})}.
\label{eq:trace}
\enq        
The local tensor $T^x$ has $2D$ indices. The explicit form is  
\begin{eqnarray}
\label{eq:tensor}
T^x_{(n_{x-\hat{1},1}, n_{x,1},\dots,n_{x-\hat{D},D},n_{x,D})}&=&\\ \nonumber
\sqrt{t_{n_{x-\hat{1},1}} t_{n_{x,1}},\dots,t_{n_{x-\hat{D},D}}t_{n_{x,D}}  }&\times&\delta_{n_{x,out},n_{x,in}},
\end{eqnarray}      
with 
the definitions
\begin{eqnarray}
\nonumber
t_n&\equiv& I_n(\beta)/I_0(\beta)\\ 
\label{eq:defs}
n_{x,in}&\equiv&\sum_in_{x-\hat{i},i} \\ \nonumber n_{x,out}&\equiv&\sum_in_{x,i}, 
\end{eqnarray}
where the sums over $i$ run from 1 to $D$. 
The Kronecker delta in Eq. (\ref{eq:tensor})
\beq
\sum_i(n_{x,i}-n_{x-\hat{i},i})=0,
\label{eq:noether}
\enq
is a discrete version of Noether current conservation Eq. (\ref{eq:cont})
if we interpret the $n_{x,i}$ with $i<D$ as spatial current densities and $n_{x,D}$ as a charge density.

The insertion of various ${\rm e}^{in_Q\varphi_x}$  is required in order to calculate the averages function of Eq. (\ref{eq:insert}). This can be done by inserting an ``impure" tensor instead of the usual one 
at the location $x$. This tensor only differs from the ``pure" tensor of Eq. (\ref{eq:tensor}) by the Kronecker symbol replacement 
\beq
\delta_{n_{x,out},n_{x,in}} \rightarrow \delta_{n_{x,out},n_{x,in}+n_Q}.
\label{eq:impure}
\enq

In Eq. (\ref{eq:trace}), the trace is a sum over all the link indices. We need to specify the boundary conditions. 
Periodic boundary conditions (PBC) allow us to keep a discrete translational invariance. As a consequence the tensors themselves are translation invariant and 
assembled in the same way at every site. 
Open boundary conditions (OBC) can also be implemented by introducing new tensors that can be placed at the boundary. Their construction is similar to the tensors in the bulk. The only difference is that 
there  are some links which could be attached at sites on the boundary and are missing. With the normalization introduced in  Eq. (\ref{eq:tensor}) the indices carrying a zero index carry a unit weight and we can 
take into account the missing links at the boundary by setting their corresponding indices to zero. 

At finite $\beta$, the ratios of Bessel functions $t_n$ defined in Eq. (\ref{eq:defs}) decay rapidly with $n$ and it is justified to introduce a truncation. If any of the indices in a tensor element is larger in magnitude than 
a certain value $n_{max}$, we approximate the tensor by zero. The main question addressed here is to decide if this type of truncation is compatible with the 
symmetries. 
\subsection{Microscopic explanation of the selection rule}
In this subsection, we provide a microscopic derivation of the selection rule Eq. (\ref{eq:selection}). 
In absence of insertions of ${\rm e}^{in_Q\varphi_x}$, the Kronecker delta at the sites can be interpreted as a divergence-free condition.
If we enclose a site $x$ in a small $D$-dimensional cube, the sum of indices corresponding to positive directions ($n_{x,out}$) is the same as the 
sum of indices corresponding to negative directions ($n_{x,in}$). 
For instance in two dimensions, the sum of the left and bottom indices equals the sum of the right and top indices. 
We can ``assemble" such elementary objects by tracing over indices corresponding to their interface and construct an arbitrary domain. Each tracing automatically cancels an in index with an out index and consequently, at  the boundary of the domain, the sum of the in indices  remains the same as the sum of the out indices.  

We can now repeat this procedure with insertions of ${\rm e}^{in_Q\varphi_x}$. Each insertion adds $n_Q$, which can be positive or negative, to the sum of the out indices. We can apply this bookkeeping on an existing
tensor configuration until we have gathered all the insertions and we reach the boundary of the system. For PBC, this means that all the in and out indices get traced in pairs at the boundary. This is only possible if the sum of the inserted charges is zero. Eq. (\ref{eq:selection}) tells us that when it is not the case, the average is zero. For OBC, all the boundary indices are zero and the same conclusions apply.

In summary we have shown that the selection rule in Eq. (\ref{eq:selection}) is a consequence of the Kronecker delta appearing in the tensor 
 and is independent of the particular values taken by the tensors. So if we set some of the tensor elements to zero as we do in a truncation, this does not affect the 
selection rule.

\subsection{Hamiltonian formulation}

The transition from the Lagrangian formulation considered above, to the quantum Hamiltonian formulation can be achieved by using the transfer matrix. 
As shown in Ref. \cite{pra90}, the transfer matrix can be constructed by taking all the tensors on a time slice and tracing over the spatial indices. 
With either PBC or OBC, there is no flow of indices in the spatial directions. Consequently the sum of the time indices going in the time slice equals the 
sum of the indices going out. This conserved quantity can be identified as the charge of the initial or final state and the transfer matrix commutes with the charge 
operator which counts the sum of the in or out indices. Consequently, setting some matrix elements to zero if some of the local indices exceeds some value $n_{max}$ in absolute value will not affect this property. The transfer matrix can be used to define an Hamiltonian by taking an anisotropic limit where $\beta$ 
becomes large on time links and the Hamiltonian will inherit the properties of the transfer matrix. 

In the rest of this subsection, we restrict the discussion to 
$D=1$ where the operator formalism is transparent.  In addition we impose  periodic boundary conditions in the Euclidean time direction. 
The tensor reads 
\beq
T_{n_x,n_{x-1}}=t_{n_x}(\beta)\delta_{n_x,n_{x-1}},
\enq
and represents the diagonal transfer matrix. In the limit of large $\beta$, $t_n(\beta)\simeq 1-n^2/2\beta$ and if we identify the time lattice spacing with $1/\beta$, 
we find the rotor spectrum with energies $E_n=n^2/2$. The value of the conserved charge $n$ is often called the angular momentum of the rotor. 
For periodic boundary conditions, the partition function is the trace of the 
$N_\tau$ power of the transfer matrix. If we insert ${\rm e}^{i\varphi_x}$ in the functional integral, the charge $n$ increases by 1 and the trace is zero unless 
we insert ${\rm e}^{-i\varphi_{x+y}}$ or a product of having the same effect. So for $D=1$, the selection rule Eq. (\ref{eq:selection}) is immediate. 
For visualization purpose, the transfer matrix evolves an initial state which is placed on the right of the operator as a ket vector and the left indices refer to the future. 

In the Hamiltonian formalism, we introduce the angular momentum eigenstates which are also energy eigenstates
\begin{eqnarray}
\hat{L}\ket{n}&=&n\ket{n},\\ \nonumber
\hat{H}\ket{n}&=&\frac{n^2}{2}\ket{n}.
\end{eqnarray}
We assume that $n$ can take any integer value from $-\infty$ to $+\infty$. 
As $\hat{H}=(1/2)\hat{L}^2$, it is obvious that 
\beq
[\hat{L},\hat{H}]=0.
\label{eq:commute}
\enq
The insertion of ${\rm e}^{i\varphi_x}$  in the path integral, translates into an operator $\widehat{{\rm e}^{i\varphi}}$ which raises the charge as in Eq. (\ref{eq:impure})
\def\cre{\widehat{{\rm e}^{i\varphi}}}
\beq
\cre\ket{n}=\ket{n+1},
\enq
while its Hermitean conjugate lowers it
\beq
(\cre)^\dagger\ket{n}=\ket{n-1}.
\enq

This implies the commutation relations
\beq
[L,\cre]=\cre, \ 
[L,\cre^\dagger ]=-\cre^\dagger, 
\label{eq:eu1}
\enq
and
\beq 
[\cre,\cre^\dagger ]=0.
\label{eq:zero}
\enq
\def\nmax{n_{max}}

We now discuss the effect of a truncation on these algebraic results. 
By truncation we mean that there exists some $\nmax$ for which
\beq
\cre\ket{\nmax}=0, {\rm and}\ 
(\cre)^\dagger\ket{-\nmax}=0.
\enq

If we now study the commutation relation with this restriction, we see that the only changes are
\begin{eqnarray}
&\ &\bra{\nmax}[\cre,\cre^\dagger ]\ket{\nmax}=1,\\ \nonumber
&\ &\bra{-\nmax}[\cre,\cre^\dagger ]\ket{-\nmax}=-1,\\ \nonumber
\end{eqnarray}
instead of 0. The important point is that the truncation does not affect the basic expression of the symmetry in Eq. (\ref{eq:commute}). 
It only affects matrix elements involving the $\cre$ operators but not in a way that contradicts charge conservation. For a related discussion of the 
algebra for the O(3) model see Ref. \cite{falko3p}.  Related deformations of the original Hamiltonian algebra appear in the quantum link formulation of 
lattice gauge theories \cite{qlink2}. It should also be noticed that Eqs. (\ref{eq:eu1}) and (\ref{eq:zero}) correspond to the $M(2)$ algebra, the rotations and translations in a plane. 
Its representations are infinite dimensional with matrix elements given in terms of Bessel functions \cite{vilenkinspecial}.

\section{Example 2: the compact Abelian Higgs model}
\label{sec:cahm}

\subsection{The model and its symmetries}

Having shown that the truncation preserve the symmetries of the O(2) model, we now proceed to discuss the question in its gauged version, the ``compact 
Abelian Higgs model". By ``compact" we mean that both the gauge field and the matter field are compact fields. On the matter side, the Brout-Englert-Higgs mode has been decoupled and the Nambu-Goldstone mode is $\varphi_x$ as in the O(2) model. 
For more details about the decoupling of the Brout-Englert-Higgs field see Ref. \cite{ahm}. The gauge fields are located on the links and are denoted $A_{x,\hat{i}}$.
The integration measure becomes 
\beq
\int {\mathcal D}\Phi =\prod_x\int_{-\pi}^{\pi}\frac{d\varphi _x}{2\pi}\prod_{x,i}\int_{-\pi}^{\pi}\frac{dA_{x,i}}{2\pi}.
\label{eq:gaugemeasure}
\enq
The action splits into a matter part
\beq
\label{eq:smatter}
S_{matter}[\Phi]=-\beta \sum\limits_{x,i} \cos(\varphi_{x+\hat{i}}-\varphi_x+A_{x,i}),\enq
and a gauge part
\beq
S_{gauge}=-\beta_p\sum_{x,i<j} \cos(A_{x,i}+A_{x+\hat{i},j}-A_{x+\hat{i}+\hat{j},i}-A_{x,j}).
\label{eq:gauge}
\enq
The symmetry of the $O(2)$ model becomes local
\beq
\varphi_x'=\varphi_x+\alpha_x
\label{eq:varphix}
\enq
and these local changes in $S_{matter}$ are compensated by the gauge field changes
\beq
A_{x,i}'=A_{x,i}-(\alpha_{x+\hat{i}}-\alpha_x),
\enq
which also leave $S_{gauge}$ invariant. The measure in Eq. (\ref{eq:gaugemeasure}) is invariant under these local shifts.

The general consequence of symmetries expressed by Eq. (\ref{eq:inv}) can again be applied to Fourier modes. We find that for {\it every} site $x$, if we have indices such that 
\beq
n+\sum_i m_i -\sum_i \tilde{m}_i \neq0,\enq then
\beq
\langle \exp(i(n\varphi_x+\sum_i m_i A_{x,i}+\sum_i \tilde{m}_iA_{x-\hat{i},i})\rangle=0.
\label{eq:gaugeselection}
\enq
This is nothing but the statement that non gauge-invariant observables have a zero expectation value. By applying this restriction to every site, we 
end up with observables such as Wilson loops or Wilson lines attached to suitable powers of ${\rm e}^{i\varphi_x}$. Even though we might not want to calculate 
the average of non gauge-invariant observable, it is legitimate to ask if truncations could generate non-zero average values for gauge-variant observables.  

\subsection{Tensor formulation}
\label{subsec:tensor}
The tensor formulation of this model has been discussed extensively in Ref. \cite{ahm} and used to propose cold atom simulations for the model
\cite{prl121}. In the following we focus on aspects relevant to a possible symmetry breaking. In order to calculate the partition function, we expand all the Boltzmann weights using Eq. (\ref{eq:fou}) and keeping the fields {\it with exactly the same signs} as in the 
cosine functions in the action. This introduces discrete quantum numbers $n_{x,i}$ for the links, just the same as for O(2), and additional quantum numbers
 $m_{x,i,j}$ associated with the plaquette with corners $(x,x+\hat{i},x+\hat{i}+\hat{j},x+\hat{j})$ and $i<j$. Comparing with Eq. (\ref{eq:gauge}), we 
 see that the gauge fields on the lowest numbered positive direction coming out of $x$ come with a positive sign and those with the largest numbered positive direction with a minus sign.   We now integrate over the gauge fields.  If we use the convention 
 \beq m_{x,i,j}=-m_{x,j,i}, 
 \label{eq:sign}
 \enq 
 when $i>j$ and in addition $m_{x,i,i}$=0, then it is clear that
 \beq
 \sum_{i,j}=m_{x,i,j}=0.
 \label{eq:sumzero}
 \enq
 We can write the  selection rules in a very compact way:
 \beq
 n_{x,i}=\sum_{j}(m_{x,j,i}-m_{x-\hat{j},j,i}).
 \label{eq:dmufmunu}
 \enq
 If we plug this relation in $\sum_i(n_{x,i}-n_{x-\hat{i},i})$, it is automatically zero because of Eq. (\ref{eq:sumzero}) and we recover the discrete version of Noether current conservation for the O(2) model. This is a discrete version of $\partial_\mu\partial_\nu F^{\mu\nu}=0$.
 
 Eq. (\ref{eq:dmufmunu}) shows that the quantum numbers associated with the links ($n_{x,i}$) are completely determined by the quantum numbers of the plaquettes 
 ($m_{x,i,j}$) which play the role of dual variables \cite{RevModPhys.52.453} but with additional  interactions given by $S_{gauge}$. The states of the Hilbert space for the transfer matrix and the associated Hamiltonian when we take the time continuum limit depend only on the $m_{x,i,j}$.
 
 So far we have only performed the integration over the gauge fields. However, the matter field $\phi_x$ appears in exponentials multiplied by $\sum_i(n_{x,i}-n_{x-\hat{i},i})$ which we just argued is zero because of Eq. (\ref{eq:dmufmunu}). Consequently, the integration over the matter fields is trivial and produces a factor 1. 
 Note that we did not fix the gauge and that the procedure is manifestly gauge invariant. The fact that the matter fields play no role here can be interpreted as a consequence of the fact that 
 they can eliminated from the action by a gauge transformation, but we did not fix the gauge. 
 
 \subsection{Interpretation of the selection rule}
 
 In the case of the global symmetry previously discussed, we found that if the sum of the inserted charges in  the full $D$-dimensional space-time volume is non zero, then  there is a flow at the boundary clashing with PBC or OBC and the average can only be zero. In the case of the local symmetry, the selection rule is microscopic and applies to a unit $D$-dimensional cube enclosing any site. 
 
 The reason gauge-variant expressions are zero is simple. For instance, it is easy to show that 
 \beq
 \langle{\rm e}^{i\varphi_x}\rangle=0,\enq
 in agreement with Elitzur's theorem \cite{elitzurt}.
 We proceed as before and integrate over the gauge fields, and all the $\varphi$'s except for $\varphi_x$. 
 If we now insert ${\rm e}^{i\varphi_x}$ in the functional integral, this is the only part that contains $\varphi_x$  since we just explained that other dependence on $\varphi_x$ disappears 
 and the integration over $\varphi_x$ produces 0 in agreement with Eq. (\ref{eq:gaugeselection}). In order to cancel ${\rm e}^{i\varphi_x}$, we need to insert another contribution,  
 for instance ${\rm e}^{-i(A_{x,1}+\varphi_{x+\hat{1}})}$, which allows us to escape the consequences of Eq. (\ref{eq:gaugeselection}) at $x$ and $x+\hat{1}$. This modifies the gauge integration and introduces non-zero values for $\sum_i(n_{x,i}-n_{x-\hat{i},i})$ which cancel the insertions of $\varphi_{x}$ and $\varphi_{x+\hat{1}}$.
 
 This mechanism persists after truncation of the Hilbert space parametrized in terms of the $m_{x,i,j}$: Eq. (\ref{eq:dmufmunu}) and its consequence 
 that we just discussed remain valid for a restricted set of $m_{x,i,j}$.
 Numerical studies of truncations in Lagrangian and Hamiltonian forms can be found in Refs. \cite{prl121,prd98}.
 \section{Extensions of the results}
 \label{sec:ext}
 \subsection{Discrete symmetries}
 The results presented in Secs. \ref{sec:o2} and \ref{sec:cahm} extend easily to the case of discrete Abelian symmetries like $Z_n$ where the shifts $\alpha$  in Eqs. (\ref{eq:phivar}) and (\ref{eq:varphix}) are restricted to integer multiples of $2\pi/n$. With that restriction, some product of Fourier modes that must have a zero expectation value for the full $U(1)$ symmetry may become non-zero if the sum of the Fourier mode vanishes modulo $n$. In a similar way, the Kronecker deltas apply modulo $n$.
 
 More generally, we never used infinitesimal transformations and as explained in Sec. \ref{sec:sym}, the measure and the action are invariant under the entire group of symmetry. The main difference in the treatment of discrete subgroups is that the sums are already finite in the original theory. 

 \subsection{Non-Abelian global symmetries}
 
For the O(3) model, the Fourier modes are replaced by spherical harmonics. For a specific global rotation $R$, Eq. (\ref{eq:insert})  becomes
\begin{eqnarray}
\label{eq:inserto3}\nonumber
&\langle&  Y_{\ell_1 m_1}(\theta_{{x}_1},\varphi_{{x}_1})\dots Y_{\ell_N m_N}(\theta_{{x}_N},\varphi_{{x}_N}) \rangle =D^{\ell_1}_{m_1m_1'} (R)
\\ \nonumber
& &
\dots D^{\ell_N}_{m_N m_N'} (R)\langle Y_{\ell_1 m_1'}(\theta_{{x}_1},\varphi_{{x}_1})\dots Y_{\ell_N m_N'}(\theta_{{x}_N},\varphi_{{x}_N})\rangle, \nonumber 
\end{eqnarray}
where the $D^{\ell}_{m m'} (R)$ are the matrices corresponding to the $\ell$ representation and the $m_i'$ indices are summed from $-\ell_i$ to $\ell_i$. By using iteratively the Clebsch-Gordan series, the
expectation value can be decomposed into a sum of irreducible representations, and only the singlets are allowed to get a non-zero expectation value.

Arbitrary truncations are likely to generate non-zero expectation values for the non-singlets. However,  if we keep 
irreducible representations at each link, in other words, if we keep all the $m$'s corresponding to a given $\ell\leq \ell _{max}$ , Eq. (3.12) of Ref. 
\cite{prd88} shows that the truncation in $\ell$ respects the global symmetries. 
This is because 
\beq
\sum_{m=-\ell}^\ell Y^\star_{\ell m}(\theta,\varphi) Y_{\ell m}(\theta ',\varphi '), 
\enq
is invariant under global rotations. 
It seems possible to extend the argument beyond this special example. 

 \subsection{Pure gauge Abelian models}
 
 The pure gauge $U(1)$ model can be obtained by taking the limit $\beta \rightarrow 0$ in Eq. (\ref{eq:smatter}). The $\varphi_x$ fields disappear from the action and their integration results in a factor 1. In the compact Abelian Higgs model, the link indices $n_{x,i}$ associated with the $\phi$ interactions are completely determined by the plaquette indices $m_{x,i,j}$ as shown in Eq. (\ref{eq:dmufmunu}). When we insert ${\rm e}^{iA_{x,i}}$ in the functional integral, an additional term is introduced in Eq. (\ref{eq:dmufmunu}) and it conflicts with  $\sum_i(n_{x,i}-n_{x-\hat{i},i})=0$ which is independently enforced by the $\varphi_x$  integration. Consequently,  for the compact Abelian Higgs model we have
 \beq
 \label{eq:elit2}
 \langle {\rm e}^{iA_{x,i}} \rangle=0, 
 \enq
 in agreement with Elitzur's theorem \cite{elitzurt}. 
 
 Extra work is needed in order to show that a similar equation is true in the pure gauge limit and that it is respected by truncations. This can be achieved by assembling tensors surrounding a given site $x$ in a way that is compatible with a selection rule.
Following Ref. \cite{prd88}, we use  $2D$ $A$-tensors
with $2(D-1)$ legs. Each $A$-tensor is associated with a link coming out of the site $x$ and its legs are orthogonal to this link. 
We assemble these $A$-tensors by connecting them with $B$-tensors in the middle of the plaquettes attached to $x$. 
Geometrically, the $A$-tensors form the boundaries of a $D$-dimensional cube. 
Graphical representations can be found in Ref. \cite{prd88}. The $A$-tensors provide a Kronecker delta that 
is a discrete version of $\partial _\mu F^{\mu \nu}=0$. 
It is expressed with a specific sign convention in Eq. (\ref{eq:dmufmunu}) with $n_{x,i}=0$.
The weight $I_m(\beta_{pl})$ appearing in the Fourier expansion of the Boltzmann weights  of the plaquette interactions can be moved to the $B$-tensor and plays no role in the discussion.

We can now imitate the procedure of Sec. \ref{sec:o2} and assign ``in'' and ``out" qualities to the legs of the $A$-tensors. For a given pair of directions $i$ and $j$, there are 8 types of legs for the $A$-tensors that we label $[(x,i),\pm \hat{j}]$, $[(x-\hat{i},i),\pm \hat{j}]$, $[(x,j),\pm \hat{i}]$, and $[(x-\hat{j},j),\pm \hat{i}]$. 
The pair of indices appearing first refers to the links where the $A$-tensor is attached and the second index to the direction of the leg which can be positive or negative. The $[(x,i), \hat{j}]$ with $i<j$ are given an out assignment. There are three operations that swap in and out: changing $(x,i)$ into $(x-\hat{i},i)$, changing $\hat{j}$ into $-\hat{j}$ and interchanging $i$ and $j$. A detailed inspection shows that this assignment gives consistent in-out assignments at the 
$B$ tensors and that the assignment is compatible with the sign partition used in Eq. (\ref{eq:dmufmunu}).  Consequently, the Kronecker delta appearing at any link is independently enforced by the Kronecker deltas on the $2D-1$ other links attached to $x$ and if we 
insert ${\rm e}^{iA_{x,i}}$ the conditions become incompatible which implies Eq. (\ref{eq:elit2}). Again the argument is based on the selection rules and is independent of the specific values of the tensors for any set of allowed indices.

 \section{Conclusions}
 
In summary, we have discussed the way symmetries are implemented in TFT for two models with a continuous Abelian symmetry. 
In both cases, we found that the truncation of the tensorial sums are compatible with the general identities reflecting the symmetries. 
By approximating some of the tensors with high indices by zero, we do not break these symmetries. The only way to do that would be to introduce new tensors which explicitly break the conservations laws at the sites or links. For numerical calculations, this implies for instance that it is possible to get a zero magnetization in the symmetric phase when a symmetry breaking term is set to zero. This is illustrated in Fig. 4 of Ref. \cite{pre89}.

For the models considered here, the symmetry is encoded in Kronecker deltas build in the tensors and located at the vertices of graphs that cover either the entire space-time lattice for global symmetries, or are enclosed in a $D$-dimensional cube for local symmetries. An intuitive picture of the way the Abelian symmetries are realized can be obtained by considering the sampling of the tensor configurations that can be performed using the worm algorithm \cite{PhysRevLett.87.160601,Banerjee:2010kc,pra90,pre93}. In this sampling algorithm, the 
worm carries a discrete charge which is conserved at each vertex following the Kronecker delta prescription. Restricting options at the vertices does not conflict with the charge conservation. 

Unlike Nother's standard field theoretical construction, our construction does not rely on taking infinitesimal symmetry transformations. The character expansions require the full group.  Consequently 
everything we did applies to discrete subgroups. For global non-Abelian symmetries, the truncation must keep a certain number of irreducible representations and combine the weights in a way that is manifestly invariant before the field integrations are performed, as we showed explicitly for the O(3) sigma model. 
It seems possible to extend this construction in more general circumstances.

Fermions are more complicated, because if we try to derive equations similar to Eq. (\ref{eq:dmufmunu}), the indices 
associated to the fermions only take a finite number of values. As fermionic theories are under construction in the tensor language \cite{Shimizu:2014fsa,Shimizu:2014fsa,Takeda:2014vwa,Shimizu:2017onf,Yoshimura:2017jpk,Kadoh:2018hqq}, this is work for the future. 
The TFT formulation of the non-Abelian Higgs model has been recently discussed and used for numerical purposes \cite{juysu2}. It would be interesting to 
try to generalize the construction of Sec. \ref{sec:cahm} for $SU(2)$. 
Another question of interest would be to understand the relationship of truncated tensor methods with quantum link models \cite{qlink97,qlink2} or matrix product states \cite{Banuls:2018jag}.

The fact symmetries are preserved by truncations means that it is  advantageous to keep these symmetries exactly in numerical formulations for instance in TRG calculations. A simple example where it is possible is given in Ref. \cite{prb87} for the Ising model where sectors of different charges can be separated explicitly. In quantum computations and quantum simulations experiments, it is desirable to have formulations with a minimal numbers of local degrees of freedom compatible with the symmetries. One can than expect to recover the result characterizing the universality class in the continuum limit. 
In noisy quantum computations, symmetry breaking is expected to occur generically and mix the energy sectors. If this symmetry breaking 
represents a relevant direction of the renormalization group flows and can be varied, results for different levels of noise and different size systems could be analyzed using finite size scaling.  Alternatively, one might try to design qubits assignments  such that the mixing of the energy sectors is impossible.

 \begin{acknowledgments}
 We thank R. Brower, E. Gustafson, S. Lloyd, W. Polyzou, J. Unmuth-Yockey, and F. Verstraete for stimulating questions. 
 This work was supported in part by the U.S. Department of Energy (DOE) under Award Numbers DE-SC0010113, and DE-SC0019139.

 \end{acknowledgments}
 
%

\end{document}